\documentclass[review=true, screen]{acmart}
\renewcommand\footnotetextcopyrightpermission[1]{} 
\usepackage{booktabs} 
\usepackage{pifont}
\newcommand{\cmark}{\ding{51}}%
\newcommand{\xmark}{\ding{55}}%

\setcopyright{none}

\begin{document}
\title{Data Capture \& Analysis to Assess Impact of Carbon Credit Schemes}

\author{Matilda Rhode, Omer Rana}
\affiliation{%
 \institution{School of Computer Science \& Informatics, Cardiff University, UK}
}
\email{RhodeM@cardiff.ac.uk}

\author{Tim Edwards}
\affiliation{%
 \institution{Cardiff Business School, Cardiff University, UK}
}
\email{EdwardsTJ@cardiff.ac.uk}

%
%
%
%
%

\renewcommand{\shortauthors}{Rhode et al.}

\begin{abstract}
Data enables Non-Governmental Organisations (NGOs) to quantify the impact of their initiatives to themselves and to others. The increasing amount of data stored today can be seen as a direct consequence of the falling costs in obtaining it. Cheap data acquisition harnesses existing communications networks to collect information. Globally, more people are connected by the mobile phone network than by the Internet. We worked with Vita, a development organisation implementing green initiatives to develop an SMS-based data collection application to collect social data surrounding the impacts of their initiatives. We present our system design and lessons learned from on-the-ground testing.
\end{abstract}

%
%

%



\maketitle

\section{Introduction}

Non-Governmental Organisations (NGOs) are increasingly using data to quantify the impact of their work. Performance metrics can be used to inform an organisation's future practices and to demonstrate the impact of programs to investors and donors. Vita\cite{vita:website}, a Dublin-based development agency, launched their Green Impact Fund initiative in 2016. The Green Impact Fund is a
financially sustainable investment initiative aimed at delivering social and environmental benefits to communities in Eritrea and Ethiopia, where the effects of global warming cause increasing agricultural and economic challenges. Companies or organisations can invest in the Green Impact Fund. Invested sums are transformed into low-cost green commodities (fuel efficient stoves, solar lamps and fixing broken water pumps). The carbon saved by these initiatives is then sold on a voluntary carbon exchange market  and the investment is returned to the investors with interest. Vita presently use in-person data collection to calculate the tonnes of carbon offset by their programs. The social impact, however, of these interventions are not measured. Collecting data in-person is costly and these surveys are conducted annually, failing to capture socio-economic benefits at a more granular level. To remedy this assessment shortfall Vita realize investors want additional indicators to specify the social impact of these interventions whilst also requiring reassurances that such work is conducted responsibly when interventions and subsequent data collection are sensitive to the needs and desires of the communities involved. We describe the process, and report the lessons learned, from developing a remote data collection application in collaboration with Vita to ascertain both the opportunities and barriers this application has for capturing social impact responsibly.

Beyond the cost of enumerator time collecting data, the present system uses paper forms before being transferred into an electronic spreadsheet for analysis. This system identifies a number of places for human error in transcription. The application also presents an opportunity to expand current data collection from the basic measurement of carbon off sets to include new lines of questioning suitable for assessing a broader spectrum of ``social impact" indicators.

Remote data collection mitigates a number of the costs of in-person analogue data collection. Harnessing existing networks incurs fewer costs than establishing or augmenting a network. Whilst just 45\% of the world have access to the Internet, there are 0.96 as many mobile phones as there are people \cite{cia:world}, though many people have more than one. Data collection tools using mobile phones exist (e.g. \cite{ushahidi}, \cite{frontlineSMS}), but existing tools were either unsuited to Vita's needs or required a developer to transform an existing software framework into a fit-for-purpose and usable application.

We describe a system designed to analyse data collected over the Global System for Mobile Communications (GSM) network, capable of assimilating data collected from individual participants and by trusted enumerators. The application for creating and sending questionnaires as well as visualising data is web-based, but transitions to operate over the GSM network for asking questions and receiving the responses. We describe the requirements gathered through interviews with Vita and other experienced enumerators, the system design, the lessons learned from user experience testing and from deploying a prototype in Eritrea.

During development we aimed to create a flexible application that might be re-purposed by other organisations. The application is intended to be usable out of the box, and should not require any programming knowledge to be used. In section \ref{related_work} we discuss how our approach compares with existing open-source tools for remote data collection. The system architecture is outlined in section~\ref{sec:architecture}. In section~\ref{sec:usage}, we discuss the lessons learned from testing our system within the context of Vita's use case, and the challenges faced in this process.

\section{Context}
\label{sec:context}

Vita, a development agency based in Dublin, Ireland, has been working in East Africa for over twenty years \cite{vita:website}. Vita launched the Green Impact Fund \cite{vita:GIF} in 2016, which seeks to use investment, rather than donation, to bring about green socio-economic improvements for individuals living in rural Eritrea and Ethiopia.

The Green Impact fund comprises several stages. Investors' money is used to repair (broken) water hand pumps and provide fuel-efficient cook stoves and solar lamps to households \cite{vita:GIF}. Each of these have both an environmental and a social benefit. Provision of clean water reduces the amount of energy (firewood) needed for sterilisation. Fuel-efficient cook stoves replace open fire pits and require less firewood for the same amount of cooking and use a chimney to extract the harmful emissions away from the household. Reduced firewood consumption may in turn reduce deforestation, which has seen around 20\% of woodland in Eritrea disappear since the 1970s for which human activity has been cited as a cause \cite{ghebrezgabher2016extracting}. Deforestation in turn has the capacity to worsen the effects of climate change and of natural disasters \cite{keller1992drought}. The solar lamps enable activities such as studying in the household after dark and may replace kerosene lamps which cause respiratory problems.

The reduction in carbon emissions from each of these initiatives can be monetised through carbon markets, with each tonne of carbon no-longer being emitted into the atmosphere commanding a \emph{market value}. Vita must demonstrate that these emissions are being saved to the independently-recognised organisation, \textit{Gold Standard}. These measurements are carried out by an external auditor, \texttt{co2balance}, using in-person data collection. In-person collection is used to ensure that certain numerical thresholds are met and that the initiatives being cited to support carbon reductions have in fact been installed. Vita wants also to demonstrate the social impact of their work, but annual, in-person data collection is costly in human hours and may fail to reflect changes that happen during the course of the year e.g. the seasonal trends in social make-up and and economic demands of the household.

Our work must also be understood in the context of Vita's efforts to demonstrate to investors that it is operating ethically, and so early work on the application is framed by the concept of responsible innovation. This concept alerts us to problems created when innovations emerge in the absence of agreed governance structures or rules to moderate the actions of researchers (\cite{wynne2002}, \cite{stilgoe2013}). Being responsible entails ``public dialogue" \cite{irwin2006} allowing open-ended governance processes to encourage deliberation. Responsibility means, ``taking care of the future through collective stewardship of science and innovation in the present" \cite{stilgoe2013}. Rather than develop the application ``at distance" from the users and communities benefitting from Vita's work, the intention is to develop this application through dialogue with communities. This enables us to make a more informed assessment of social impact of this work, and the use of the application can be modified in ways responsive to the realities of the respondents' experiences and needs. This is especially important in a context where the success of Vita's operations rely on inclusivity and when the Eritrean authorities are highly suspicious of possible foreign interference in national matters.

The table below indicates the summary statistics for Ethiopia and Eritrea according to the CIA World Factbook estimates\footnote{Data estimates from July 2015}\cite{cia:Eritrea, cia:Ethiopia, cia:world}

\begin{table}[h]
\centering
 \begin{tabular}{||c | c c c||}
 \hline
total / population & Ethiopia \cite{cia:Ethiopia} & Eritrea \cite{cia:Eritrea} & World \cite{cia:world}\\
 \hline\hline
Landlined phones & 0.01 & 0.01 & 0.15 \\
 \hline
 Mobile phones & 0.41 & 0.07 & 0.96\\
 \hline
 Internet Users & 0.12 & 0.01 & 0.43\\
 \hline
\end{tabular}
\caption{Ratio of landlined phones, mobile phones, and internet users to total population in Ethiopia, Eritrea, and globally}
\label{population_communication}
\end{table}

The data above indicates the relative numbers of mobile phone ownership to population size. Globally less than 96\% of the population are connected to a mobile phone network -- as some individuals have multiple phones. Whilst mobile phone ownership is relatively low in Ethiopia and Eritrea compared with global rates, it it still significantly higher than the number of individuals with internet access. Furthermore, smartphone ownership rates are lower than basic phone ownership at 8\% in Ethopia \cite{Pew15}, compared with 43\% owning a basic or featureless phone. In Eritrea third generation (3G) mobile wireless mobile telecommunications technology is not available \cite{GSMArena}. Hypothesising that mobile phone ownership is positively correlated with wealth, we conjecture that direct data collection from individuals will not conduct a balanced survey for this use case.

We selected SMS communication above voice communication after researching similar ventures in other developing countries. Whilst \cite{patnaik} found that voice reporting gave fewer data errors than SMS reporting for collecting health data (in Gujarat, India), participants were trained prior to using the software. Crawford et al. \cite{crawford} argue that SMS is the preferred means for disseminating health data. This is due to lower costs, higher delivery success, and higher levels of intended or actual behaviour change. The context of\cite{crawford} matches the environment of the Green Impact Fund more closely, involving untrained voluntary participants communicating over an unreliable network infrastructure. Voice-based communication, however is more accessible to a wider population. Eritrea has a literacy rate of 73.8\% \cite{cia:Eritrea}, and Ethiopia 49.1\% \cite{cia:Ethiopia}. Furthermore, the Green Impact Fund primarily focuses on providing their fuel efficient cook stoves and solar lamps to women. In both countries, the literacy rate among women is lower than the countrywide average and in Ethiopia; women are also less likely to own a phone \cite{unesco15}. Despite this, network reliability can interfere with voice transmission or potentially cut phone-calls short. The former may increase transcription errors (for data analysis) and the latter introduce a bias towards the initial questions in a survey having a higher response rate.

To combat the partial and biased ownership of mobile phones as well as barriers to text communication, we build in the flexibility for trusted enumerators to collect data as well as for individuals to provide data about themselves. This is still of value to organisations as it allows dynamic changes to data collection (from a web application) in the field and also benefits from the automated analysis outlined in section~\ref{subsec:parsing}.

\section{Related Work}
\label{related_work}

Remote data collection is attractive for its potential to reduce both the time and financial cost of in-person collection. Existing tools for collecting data vary according to their purpose. Tools may be developed for specific projects or may be flexible enough to meet the needs of a number of different projects. The suitability of a given tool to a project may be restricted by the type and capability of telecommunication network(s) over which data is collected, the intended users, the possible data formats etc. The system we present in section~\ref{sec:architecture} seeks to add to the set of existing tools for remote data collection with reference to Vita's specific use case, and others like it.

\subsection{Remote data collection tools}
\label{sec:remotedatacoll}

Applications designed for a specific purpose may aggregated data to identify trends, e.g. information about crises \cite{ushahidi}, or singular data points may be useful, e.g. for reporting road incidents in Nairobi \cite{communisense}. Communication channels can also be reversed to disseminate data, e.g. health information \cite{crawford} or agricultural information \cite{Qiang}. It is also possible to combine the two by disseminating collected information. Although tailoring tools to specific use cases may enable greater efficiencies for the project in question, Tangmunarunkit et al. \cite{Ohmage} highlight the inefficiencies of this approach when many projects have very similar requirements. Other tools are flexible and have been designed to fit the needs of various remote data collection objectives (\cite{brunette:odk2}, \cite{frontlineSMS}, \cite{Ohmage}, \cite{epiCollect}). Open source applications such as the Open Data Kit (ODK) \cite{brunette:odk2} further allow developers to tailor flexible applications to specific projects or to develop software packages to add greater flexibility to the application. Some tools, such as RapidSMS \cite{rapidSMS} are specifically designed as a framework upon which developers can build, the disadvantage here being that a developer is required to create a usable application. Flexibility is maximised when applications can be used out-of-the-box and also developed further by software engineers if required. Both Ohmage \cite{Ohmage} and ODK fulfil these criteria, but require an Internet connection to send data.

Polling and surveying utilises networks, whether these be terrestrial or digital. Internet-based solutions such as the Android-based ODK and iOS/Android-compatible Ohmage and EpiCollect have several advantages over collection applications using the GSM network. Firstly, multimedia data can be sent easily over the Internet, e.g. images and geolocation data. Secondly, Internet-based collection can reduce costs by storing data locally on enumerator devices until a cheap Internet source is available. SMS messages and voice calls cost per message or per minute and so savings can only easily be made through  ``unlimited" plans. However, as shown in table~\ref{population_communication}, Internet connectivity is less prevalent than GSM network connections globally.

Tools may impress a data structure onto the collection process. FrontlineSMS \cite{frontlineSMS} is highly flexible and does not impose a structure on data collection. The advantage of this approach is the flexibility afforded to the question-setter. Collecting data can be riddled with concerns over introducing biases and as such it may be undesirable for the software to place restrictions on question designers. ODK, conversely uses questionnaires and structured response formats (such as check-boxes) with data types, e.g. geolocation data. This approach can enable quick data analysis either visually or numerically with reference to the questionnaires, questions and response location. In cases of open source applications, data analysis tools may be added on more easily with structured data responses, as illustrated by Imran et al. \cite{imran2016enabling} in their analysis of health care questions submitted by SMS to U-Report Zambia \cite{ureportZambia}, to which trained health care professionals then respnd. In \cite{imran2016enabling}, questions are sent by SMS then classified into categories using a trained machine learning classifier with an accuracy of 82\% area under the curve. The authors simulated a \textit{live testing} scenario and the health workers reported that the system helped to process the large volume of SMS queries more quickly. Imran et al.~\cite{imran2016enabling} demonstrates how automatic parsing of SMS messages can elevate the usefulness of an initiative, particularly in cases where action might be time sensitive. The authors note the difficulty in parsing natural language and of categorising messages which fell into one or more of the predetermined categories. As our model is based on a system which prompts for answers, in this initial phase we simply count word frequency to get a sense of the responses from users. In future iterations of the application machine learning could be employed in a similar fashion to \cite{imran2016enabling} for collecting data on particular topics.

Whilst ODK \cite{brunette:odk2}, EpiCollect \cite{epiCollect} and Ohmage  \cite{Ohmage} comprises many of the features required by Vita, each of these applications operates using feature phones and is not appropriate for use in Eritrea.

\subsection{Data collection using SMS}
\label{sec:usingsms}

Many NGOs use SMS to disseminate and/or collect information. The success of specific projects such as UReport~\cite{ureport} to promote citizen engagement and Ushahidi \cite{ushahidi}, initially developed following the riots connected with the 2008 elections in Kenya, have now been rolled out to other countries as reporting mechanisms operating more generally. These applications aggregate data to sense changes in population reporting, but do not enable organisations to prompt answers to specific questions, as is required by Vita.

Frameworks such as FrontlineSMS and RapidSMS, commissioned by UNICEF in 2007, can be built on to create data collection applications for specific use cases. These frameworks require a developer to create the application, however. Our framework below is intended to suit Vita's needs. We hypothesise, based on the success of applications such as ODK collect, that a structured method for polling using SMS will fit many use cases as well as Vita's. In creating an SMS polling mechanism it will also be possible to disseminate data (technically equivalent to asking a question requiring no answer), thus widening the potential uses of the application.

\section{System Requirements}
\label{requirements}

In designing the system we wanted to build on the existing knowledge of experienced data collectors and conduct tests to closely approximate the situations in which the application might be used. Vita and co2balance were vital in the development of the project. The initial impetus for the system described here was borne out of Vita's Green Impact Fund launch event, during which it became apparent that potential investors were interested in quantifying the social impact of their donations.

\begin{sloppypar}
To prove compliance with the Gold Standard, Vita employ co2balance to conduct in-person data collection. Though Vita could append social impact questions to this existing data collection practice, there were concerns that this would make the questionnaire too long, increasing the potential for data bias due to respondent fatigue. Furthermore these questionnaires are carried out annually and Vita may want to collect social data at closer intervals. It is also worth noting some reasons that remote data collection is not always preferred. Firstly, as mentioned above, not every respondent has access to a mobile phone, thus data will be biased towards representing those with access. There are a number of other biases that may be introduced through remote collection, as SMS messaging is not free, there may be a bias towards wealthier respondents. There may also be a bias towards those who are more frequent users of mobile phones, however it is also possible to argue that door-to-door surveys are biased towards representing the views of those most likely to answer the door. This system is not intended as a panacea for all issues surrounding in-person data collection, but as a complementary tool. As such it can be used both by enumerators and by individual participants, with the system aggregating responses to the same question regardless of how it was collected. Responses collected by paper should also be possible to add to the system in case Vita wishes to incorporate old data into the new system.
\end{sloppypar}

Our research repeatedly indicated features which represented a trade off between convenience and data integrity. If users are experienced enumerators, the system should not impede on the way in which they wish to collect data and if they are not, perhaps a superuser should be able to curb their actions within the guidelines of good data collection practices. We built on the information gathered from Vita and co2balance and took insights from papers such as \cite{cobb:computer}. In\cite{cobb:computer}, the authors interview a number of NGO employees using ODK to tease out the security concerns of enumerators, during which the authors discover that many enumerators fear data loss as well as exposure of sensitive data. We aimed to protect against the possible causes of data loss such as accidental or purposeful tampering by an authorised and unauthorised individuals as well data lost as as result of poor network integrity.

From our research we distilled the following three principles that the application should seek to meet:

\begin{enumerate}
	\item \textit{Facilitate the collection of as much (accurate) data as possible:} promote those aspects which will encourage users at both ends of the system to generate accurate data;
	\item \textit{Automate:} use computation to save resources (without violating 1)
	\item \textit{Provide flexibility:} design the system such that it might be adopted for different forms of SMS communication i.e. information dissemination, collection and collection followed by dissemination.
\end{enumerate}

\section{System architecture}\label{sec:architecture}

Using the GSM network, communication may be conducted via text or using voice communication. Whilst voice communication does not entail literacy requirements, we present a text-based application using short message service to capture the data analytic capabilities of Internet-based applications such as Open Data Kit  \cite{brunette:odk2}, using structured questionnaires. Due to our specific use-case, Vita's concerns over network robustness leant in favour of SMS communication as voice-based data collection occurs at the inconvenience of the participant and network failures may lead to a bias of more data for earlier questions in a given questionnaire. Flexibility for collection by an enumerator is the temporary solution to literacy requirements, but we chose an SMS gateway provider (Twilio~\cite{twilio}) capable of using voice communication with a view to incorporating this feature in the future.

The robustness of SMS is built into its protocol. Although SMS messages are subject to a 160 character limit, longer messages are enabled by ``concatenated SMS". Concatenated, or long SMS carry a header of metadata to indicate that they are one in a series of messages, along with their position in that series \cite{leBodic}. If all parts of a message, or even any part, is not able to be sent within the validity period for the message as specified by the network operator, typically two days \cite{leBodic}, the message will fail to send. In a questionnaire context this means the next question will not be triggered. If the gateway is unable to detect a failed message, should the next question be sent after two days if there is no response? Or did the respondent choose not to reply and would rather not be bothered?

The system presented here effectively creates a structured system for two-way communication. At one end of the system sits the data analyst and/or questionnaire architect, at the other a data enumerator or an individual sending responses to the server. To harness the mass-polling capabilities of SMS we use an SMS gateway across a Global System for Mobile Communications (GSM) network. Though the use of GSM is intended to reach areas which have poor or non-existent internet connectivity, we have designed a web-based application for the data analysis and questionnaire-setting end of the system. The web-based application suits the structure of decentralised and globally disparate organisations such as Vita, with employees in Eritrea, Ethiopia and Ireland, as data and questionnaires can be accessed and sent from anywhere with an internet connection. In Eritrea and Ethiopia, Vita's headquarters are based in the capital and will have access to Internet.

\begin{figure*}
	\includegraphics[width=0.7\textwidth]{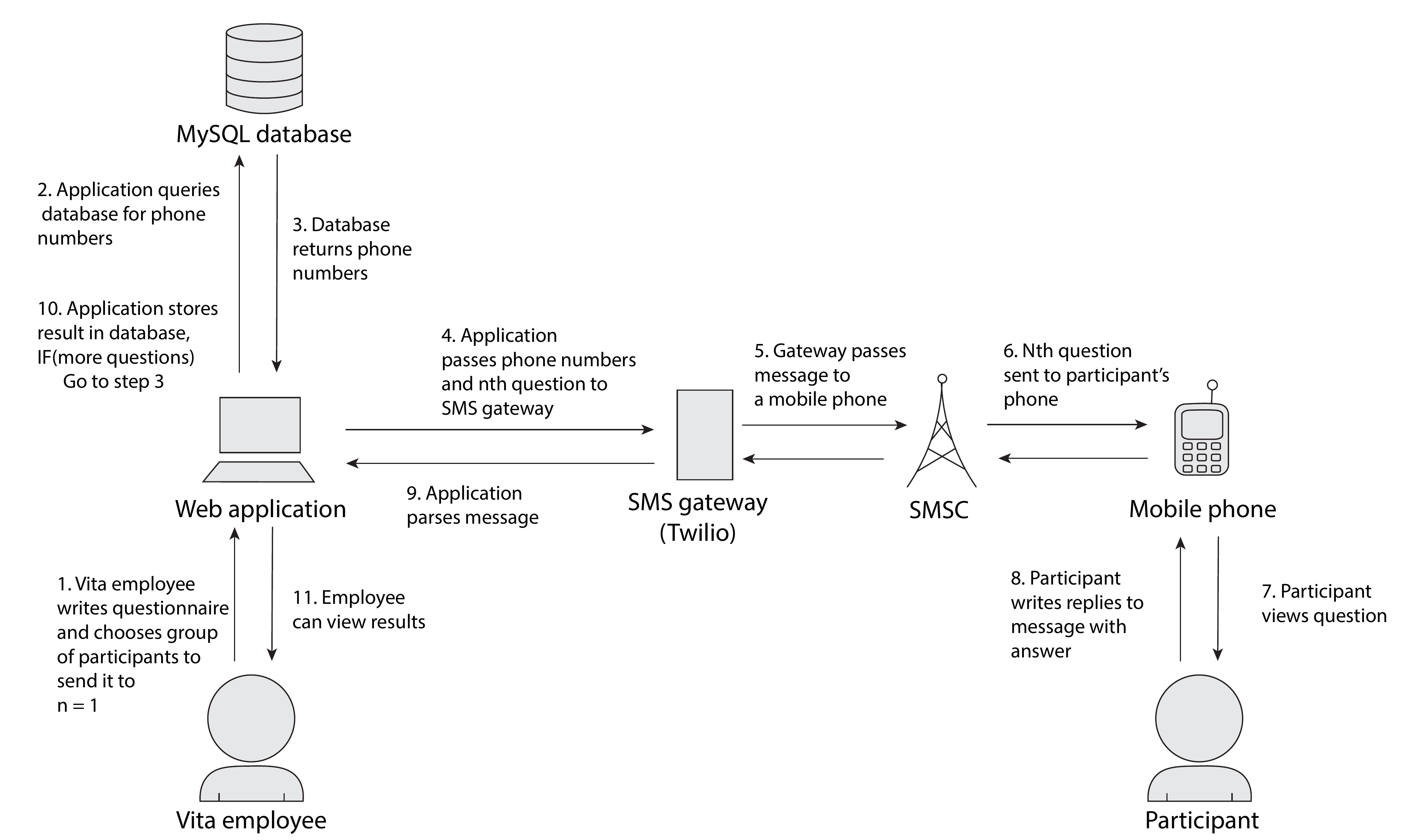}
	\caption{Data flow within application}\label{data_flow}
\end{figure*}

Figure~\ref{data_flow} illustrates the application architecture. The database houses all questions, questionnaires, responses, respondent and employee data. The SMS gateway connects the web application and a short message service center (SMSC), which then transmits the message to the intended recipient(s) over the GSM network. This model can be adapted to enable information dissemination as well as collection. The web application directly pulls and pushes data to the database, queries to the database are only made when editing data (not when changing view) thus reducing the number of requests to the database server and potential for failed requests due to unreliable Internet connections.

\begin{figure}
	\includegraphics[scale=0.25]{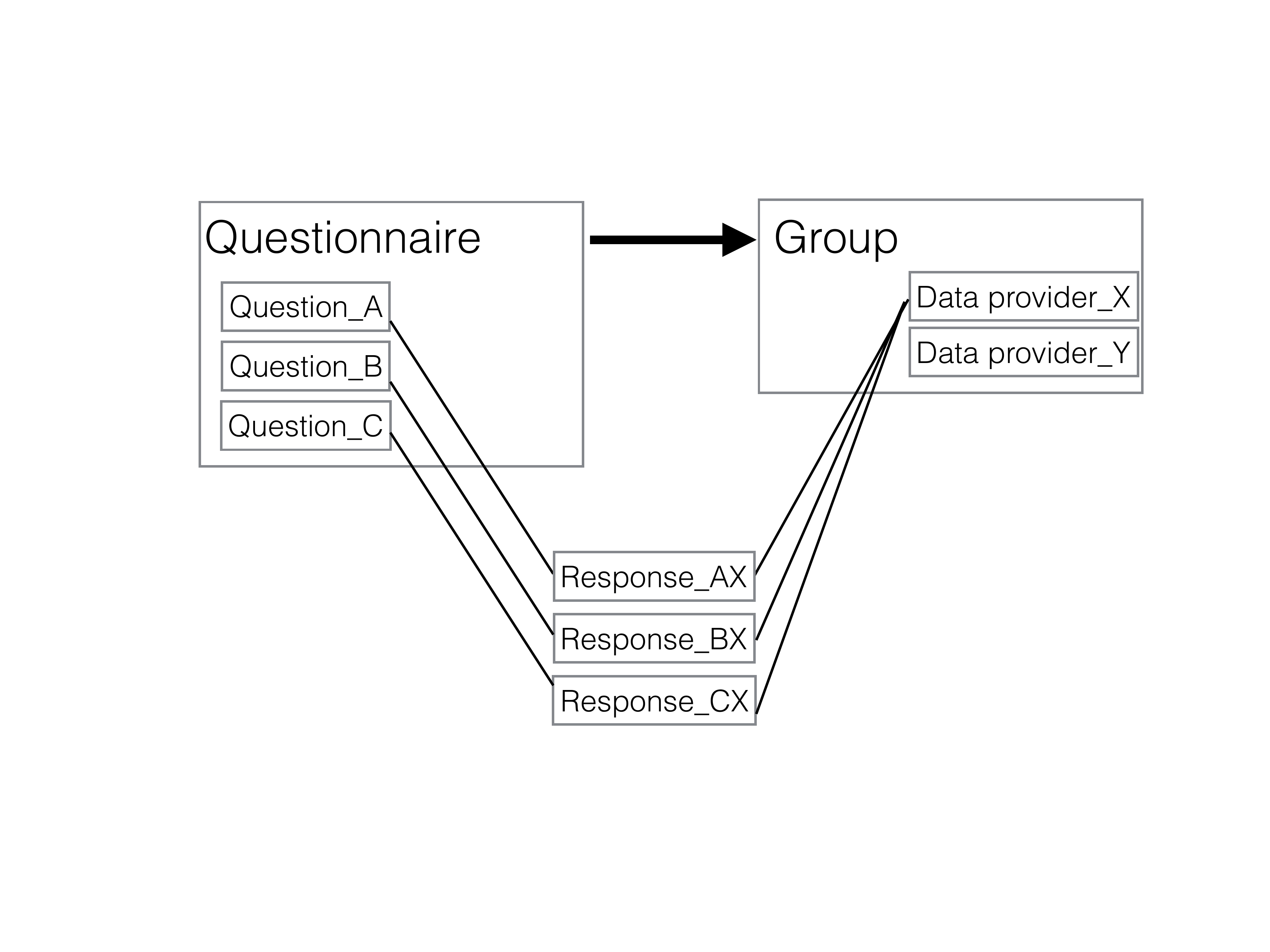}
	\caption{Illustration of questionnaire relationship to user group}
	\label{data_structure}
\end{figure}

The data structure for sending questionnaires operates using questionnaires and user groups. This enables the polling of subsets of individual respondents as well as creating a group of enumerators, who may be sent different questions to users. Figure~\ref{data_structure} gives an outline of a situation in which a questionnaire, containing three questions, has been sent to a group, containing two members. Only one member has responded and they have replied to all three questions.

\section{SMS data collection application}

\begin{figure}
	\includegraphics[width=0.5\textwidth]{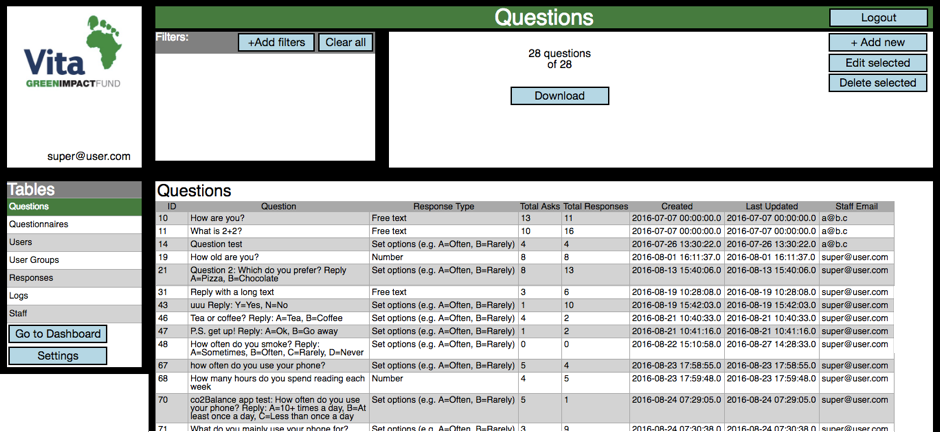}
	\caption{Web application view of questions in database}
	\label{questions_table}
\end{figure}

\begin{figure}
	\includegraphics[width=0.5\textwidth]{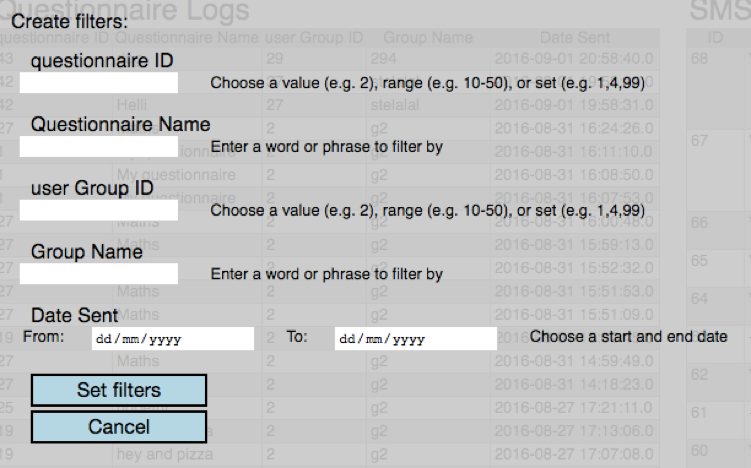}
	\caption{Filter overlay for user groups}
	\label{filter_overlay}
\end{figure}

The web application is the management portal for creating and sending questionnaires as well as for data analysis. For consistency the website employs a tabular system comprising questionnaire, user and response data, as well as staff data for superusers. A sidebar on the left hand side enables navigation between tables (Figure~\ref{questions_table}). Each table may be filtered according to its contents (Figure~\ref{filter_overlay}). Multiple filters can be used at once to show the subset of rows fulfilling all applied filters.  Filtering enables shortcuts to common processes such as creating user groups or questionnaires and downloading data as a comma separated values (CSV) file. Depending on access levels (Figure~\ref{access_levels}) employees are able to add, edit and delete table data.

\subsection{Creating and sending questionnaires}

\begin{figure}
	\includegraphics[width=0.5\textwidth]{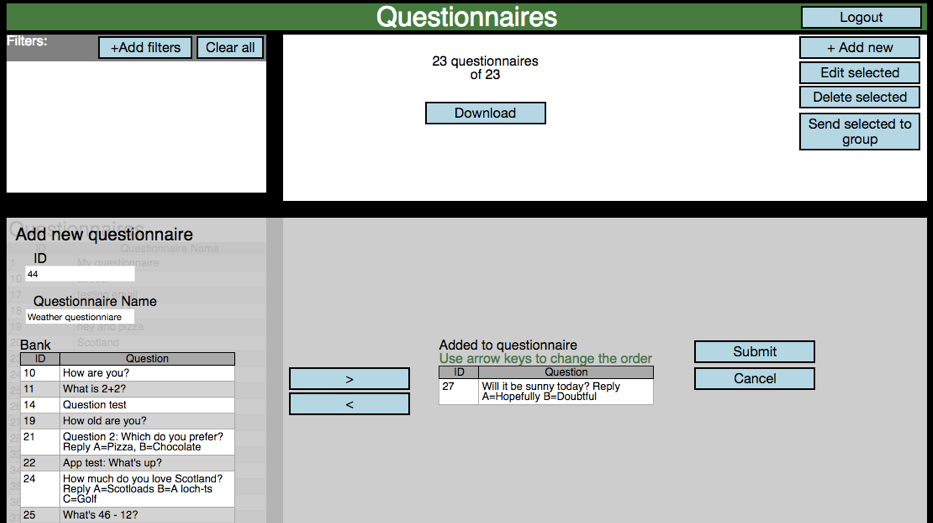}
	\caption{Creating a questionnaire}
	\label{creating_questionnaire}
\end{figure}

Creating and sending a questionnaire requires a questionnaire and a user group, comprising one or more questions or users respectively. Questions are added to questionnaires and users to user groups either by manually selecting from the list of possible candidates (Figure~\ref{creating_questionnaire}) or by filtering the questions/users table and choosing  ``Create [Questionnaire / User Group] from selected".

\begin{figure}
	\includegraphics[width=0.5\textwidth]{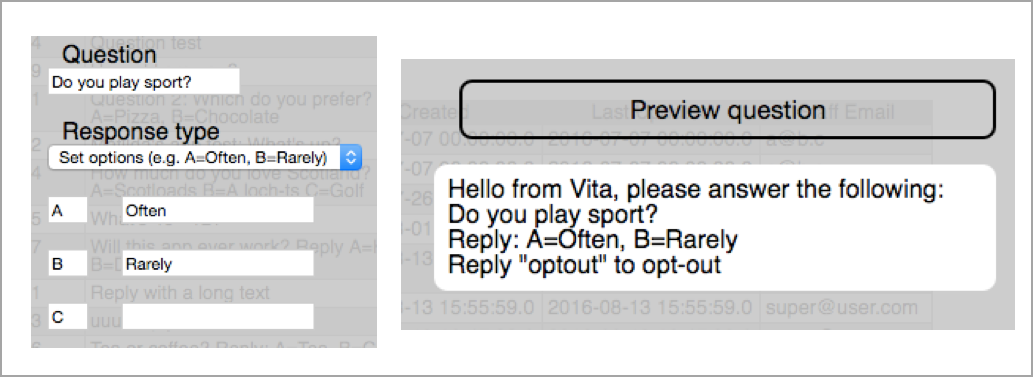}
	\caption{Previewing a question before committing it to database}
	\label{preview_question}
\end{figure}

\begin{figure}
	\includegraphics[width=0.5\textwidth]{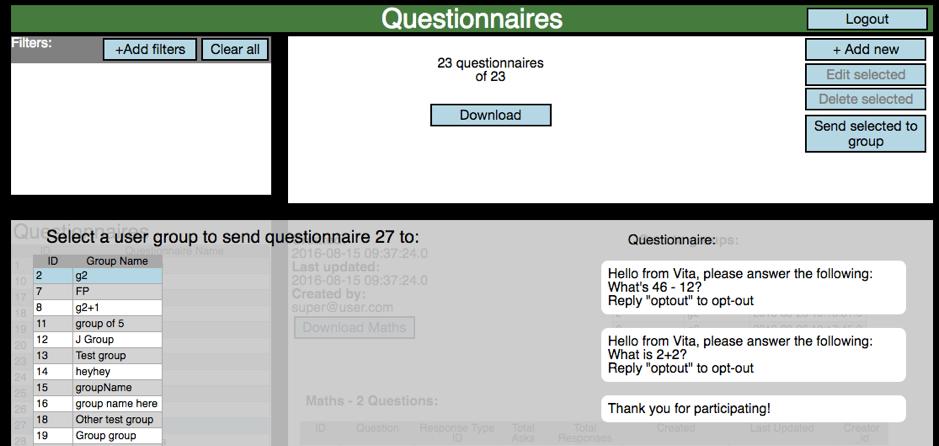}
	\caption{Preview questionnaire before sending to user group}
	\label{preview_questionnaire}
\end{figure}

Questions themselves must be given response types to enable automated response parsing and data analytics. Response format may be numerical, free text or categorical options. After user experience testing we added a  ``Yes/No" option, which though logically indistinct from a categorical response, was deemed a time-saving and rational option to our testers. Employees are restricted the formatting of options questions based on research by Down and Duke \cite{downDuke} showing the format which was easiest for respondents to understand. Maintaining format also enhances ease of use for repeat data providers. Employees may preview individual questions (Figure~\ref{preview_question}) as well as entire questionnaires (Figure~\ref{preview_questionnaire}). Previews also reveal settings put in place by the superuser(s) including greetings, thank-yous and opt-out messages.

The user table can be considered a series of  ``profiles". Certain data cannot be removed such as profile creation date, last edit, and phone number; but employees may add further columns to the tables to create more user data. For example birth date and languages spoken may be included enabling quick filtering so that a user group of all under-25s may be created and polled. This feature permits the flexibility to use enumerators rather than polling individuals. An organisation may add an attribute (column) to user profiles denoting  ``staff". Staff may then be polled in the field with per-household, per-street, per-village questionnaires as preferred.

We imagine a scenario in which an organisation's employees hold various roles. Staff profiles are each attributed access levels which enables the application to run with least privilege. This is a security measure against both tampering and accidental data loss. The access levels are denoted in Table~\ref{access_levels}.

\begin{table}[h]
	\small
	\begin{tabular}{|p{0.12\columnwidth}|p{0.07\columnwidth}|p{0.07\columnwidth}|p{0.07\columnwidth}|p{0.07\columnwidth}|p{0.07\columnwidth}|p{0.07\columnwidth}|p{0.07\columnwidth}|p{0.07\columnwidth}|}
	\hline
	Data Access Level & View & Filter & Down-load & Add & Edit and Delete & Send SMS & View \& manage staff & Update sys. settings \\ \hline
	Read only  & \cmark  & \cmark  & \cmark & \xmark& \xmark & \xmark & \xmark & \xmark  \\ \hline
	Add  & \cmark  & \cmark & \cmark  & \cmark & \xmark& \xmark& \xmark& \xmark \\ \hline
	Edit  & \cmark  & \cmark & \cmark &  \cmark & \cmark & \xmark & \xmark& \xmark \\ \hline
	Send  & \cmark  & \cmark & \cmark  & \cmark  & \cmark & \cmark & \xmark& \xmark \\ \hline
	Superuser & \cmark  & \cmark  & \cmark & \cmark  & \cmark & \cmark & \cmark & \cmark\\ \hline
	\end{tabular}
	\caption{Employee Access Levels}
	\label{access_levels}
\end{table}

Superusers may place further restrictions on employees beyond their access levels to promote good practice in data collection. For example superusers can set a maximum questionnaire length to mitigate the chances of respondents becoming bored by questionnaires, resulting in a fatigue bias. Superusers may add other features to questionnaires such as greetings and thank-you messages. Featureless phones do not present SMS messages in a  ``thread" style so the greeting will enable respondents to recognise the organisation polling them if they have not saved the relevant number or message code.

\begin{figure}
	\includegraphics[width=0.5\textwidth]{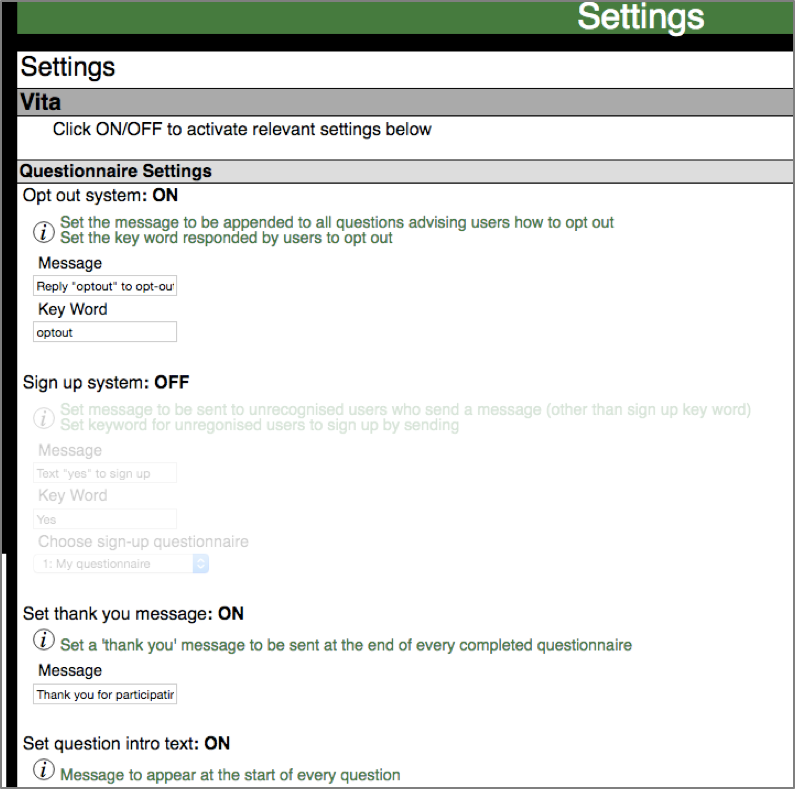}
	\caption{System settings in web application}
	\label{settings}
\end{figure}

Superusers may further choose special keywords to allow respondents to opt-out or sign-up to receiving questionnaires using SMS. The sign-up word can be associated with a questionnaire which will be sent after the sign-up keyword. The opt-out word deactivates the user profile until the user sends the sign-up keyword, this is so that their previous responses remain associated with a user profile and data is not lost. These features are enabled through the system settings, which are only accessible by superusers (Figure~\ref{settings}).

\subsection{SMS response}

Our design borrowed lessons from the models used by market research companies and others using SMS to collect data; the questionnaire waits for a response to the last message sent before sending the next question, this stops the user from being repeatedly asked questions when they have no interest in or are unable to respond. An alternative to question-response-question-response is to send the next question if no response had been received following a pre-specified time period. As the short message protocol encompasses an acknowledgement of message receipt \cite{smsTechnicalRealisation}, a lack of response can be attributed to a user deciding not to respond or network integrity issues on sending the response. The short message protocol should re-attempt sending until success or until two days elapses. Failures are likely to be due to network error, and may be deduced by noting that all respondents from a particular area have not replied to questions sent within a particular timeframe. Such an occurrence, however may also be an active decision not to respond. The web application contains a complete and uneditable log of all SMS sent and received to the gateway. This is intended for data integrity and may be used at an NGO employee's discretion to decide whether respondents' decisions or network integrity is the cause of a lack of responses.

Communicating strictly by one new question following an answer to the previous question also enables data parsing to happen automatically. A piece of data is only useful if the question to which it responds is known. It may seem over simplistic to denote an SMS as a response to the last question sent but Church and de Oliviera \cite{ChurchWhatsapp} note that the cost involved in sending SMS messages creates a more structured answer-reply conversation than occurs over instant messaging platforms, in which users will frequently send sentences as single messages in quick succession. If the application does falsely attribute a response to a question, the analyst may examine the SMS logs to determine the question for which it was a response, if any, and alter this in the responses table.

For flexibility we chose to use an SMS gateway API rather than a hardware module, which would be more difficult to distribute and require technical expertise to install. Testing revealed that although our chosen gateway communicated easily between some international network providers, others blocked responses to our tests from being sent and some blocked their delivery. This information was not always available on the API website. There were a number of possible solutions to this problem e.g. using hardware modules or negotiation with network providers.

\subsection{Parsing and analysing data}\label{subsec:parsing}

To further reduce the human resources required to process the data, we have included automated parsing of question responses to facilitate data analysis. We concede that automation of human responses may parse data incorrectly for certain edge-cases, so the system maintains a complete log of messages sent and received to stop misinterpreted data being overwritten permanently. Misinterpretation may result from a question response being sent in multiple parts, as described in the previous section or from mis-parsing of categorical data, though we believe that both of these are unlikely occurrences. Each response holds a one-to-one relationship with a user and with a question, metadata about the responses is also stored. Categorical response questions are in turn related to possible answer codes and the meaning of these codes e.g.  ``A=Everyday, B=Weekly, C=Monthly, D=Never". We use simple case-insensitive parsing on responses to categorical answer type questions. Answers are stripped of start and end spaces and punctuation to translate  ``a",  `` A." and  ``everyday", for example, to  ``Everyday". The purpose of this parsing is to enable quick data analytics. Response parsing also validates inputs to protect against the possibility of an SQL injection attack, which could cause data to be lost or corrupted.

After logging-in, the landing page for employees is a dashboard, where user and question data is displayed in simple visualisations. Data gathered from the database is transformed into a JSON format representing summary statistics for user attributes and question responses. The visualisations use the d3 JavaScript library for data visualisation \cite{d3}. A further development could easily be added to download the summary statistics as a JSON file for analysis use other software.

\begin{figure}
	\includegraphics[width=0.7\columnwidth]{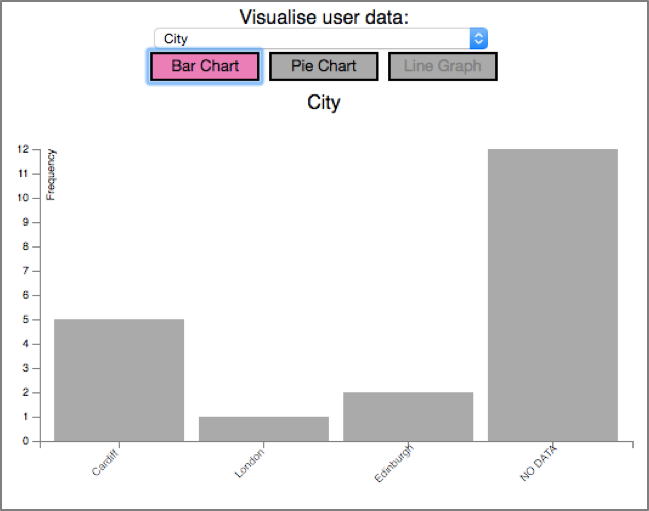}
	\caption{User city data visualised in suggested bar chart form, can be overridden to view pie chart representation}
	\label{barchart}
\end{figure}

\begin{figure}
	\includegraphics[width=\columnwidth]{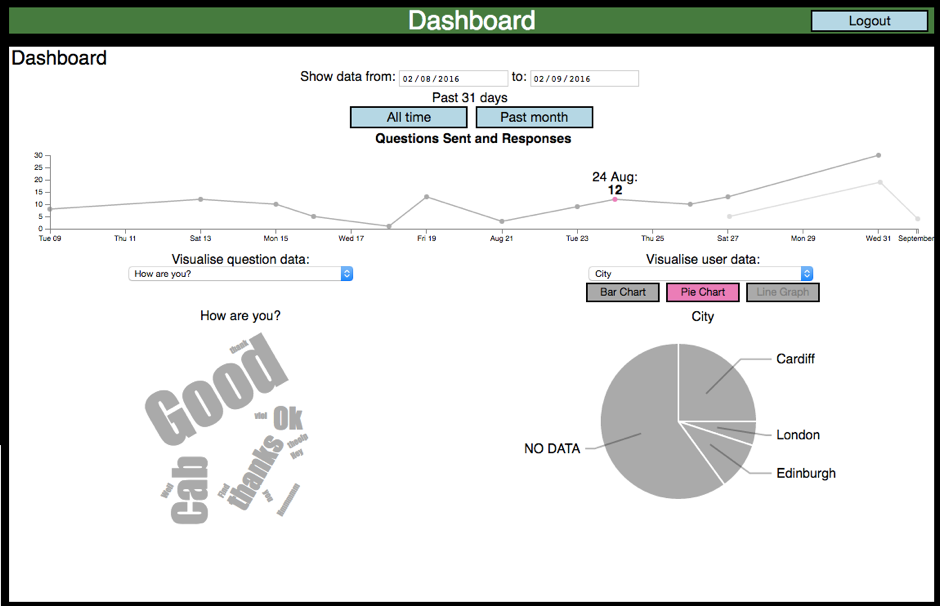}
	\caption{Screenshot of the dashboard during development}
	\label{dashboard}
\end{figure}

The dashboard comprises a timeline for an at-a-glance indiction of the number of SMS being sent and the number of responses being received. The data relating to each question can also be visualised. Employees select the question from a drop-down menu and data is displayed as a bar chart, pie chart or word cloud if the response type was numeric, categorical or free text respectively. Similarly, user attributes can be selected from a dropdown menu on the right Figure~\ref{dashboard}) and a visualisation type is suggested based on data format, though this can be overridden (Figure~\ref{barchart}). These visualisations are intended for a quick overview of demographics and responses; all data may be downloaded as a CSV file if analysts wish to conduct further investigation in another application.

\section{Implementation and user experience testing}

Interviews with experienced data enumerators and research informed the application's development but we wanted to conduct tests to find further areas for development and improvement. We conducted a user experience test for the web application, an international data collection test, and interviews and system tests with Vita in Eritrea. Four challenges stood out as barriers to the usefulness of the application: network provider blocking of SMS gateways, distribution of mobile phone access, impact of network integrity on enumerator data collection and graphical interface changes for greater ease of use.

Initially we had problems communicating internationally by SMS between the UK and some countries.  Using the SMS gateway service provided by Twilio~\cite{twilio}, we were able to communicate nationally and with several European countries but experienced difficulties conducing two-way communication in Kenya. We later adapted the system to use a GSM modem containing a SIM card and the Ozeki SMS Gateway \cite{ozekisms}, which circumvented this issue. A further solution is to negotiate with network providers to allow bulk SMS traffic through from pre-specified phone numbers. The advantage of the latter solution is that the GSM hardware acquisition and installation does not need to be performed.

The mobile phone statistics reported by the CIA world factbook for Eritrea \cite{cia:Eritrea} give a figure for the entire nation. Brief interviews in Eritrea indicated that probability of mobile phone ownership grows in proximity to Asmara. In one village close to Asmara, residents estimated that ownership stood close to 40\%. Although we did not conduct a numerical survey, this figure is much higher than 7\%. Even with a significant margin of error allowed for estimation biases, the implication is that ownership is much lower than 7\% in other areas to give this national average. This information emphasised the importance of flexibility so that enumerators could collect data and not rely on participants owning mobile phones to provide data.

We discussed improvements to the application to enable enumerators to provide data quickly and cheaply. If network speeds fall or a network failure occurs while enumerators are in the field collecting data, they do not want to wait for a text asking them the next question to reply. Furthermore the cost of sending a text message is determined per message (if the message is less than 160 characters). We began to develop an encoding system so that enumerators could send data for ten questions in one go. This relies on keywords and would necessitate some training of enumerators. The keywords need to signal questionnaire, question, and respondent. This may appear very close to the paper questionnaire model but its advantages are that human error is mitigated, data is automatically parsed for analysis and this data is integrated with the individual respondent and other enumerators' data. Enumerators would need a copy of the questionnaire, this could be stored electronically, but this would require frequent switching between applications or SMS messages in the phone, two devices or learning by-heart, so a paper copy was deemed the best alternative. The questionnaire number is printed at the top of the page and a list of user data stored too. Keywords are reduced to letters to conserve the 160-character limit. The message style uses codes and ID numbers to establish questionnaire, starting question and respondent and then uses semicolons to separate question answers e.g.  ``U24F102Q1;A;12;C;More time with family" could be the responses of user 24 in answering questionnaire (form) 102 from questions 1 to 4.

This development later inspired the ability to combine short SMS into single text messages to reduce the cost of the respondent. This feature is still under development. The hesitation over implementing this feature is in understanding the trade-off between the cost of sending an SMS as a deterrent and the data loss that will occur from human error in inputting the data (for untrained individual participants). In the future we would like to conduct experiments to assess the relative merits and demerits of these two systems. As the relative impact of these factors is likely to depend on context, the superuser will have powers to activate or deactivate this feature for flexibility.

We conducted an initial round of user experience testing (four volunteers) in which we asked users to complete various tasks after explaining the purpose of the application but not how to use it. The results were that the membership of questions to questionnaires and users to user groups should be more fluid to enact, and there should be more ways of doing this.

Plans to extend this work would ideally involve user group workshops within targeted communities to explain the application, the purpose of the study and how to use the application to enable \textit{live} data collection over time. However, for this to be done in a responsible way it would also be necessary to engage in dialogue about the fears and desires of respondents in participating in such an evaluation. Transparency is crucial, so \textit{lessons learnt} inform monitoring and evaluation protocols moving forward alongside the expansion of Vita's work planned between now and 2021. In this respect, the application might also be viewed as a mechanism to enable dialogue and not simply as a means to report impact. Indeed, the application has the potential to improve impact through knowledge dissemination.

\section{Usage Experience}
\label{sec:usage}

The system has been used in Eritrea to collect data for the Vita Green Impact Fund. Accurate data collection remains an important challenge for an NGO of this kind, and this was the key reason for developing the proposed system. We gained the following experience from this study:

\begin{itemize}

\item Although mobile phone usage and access remain high in this region, access to \textit{smart phones} with internet access  remains limited. A system that requires data collection to be supported directly based on an internet-enabled infrastructure is of limited value. The telecomms. infrastructure may be provisioned through a private company, its operation is often controlled by government. This requires any additional service which must operate over such infrastructure to go through various approval processes.

\item The energy infrastrucuture required to sustain data collection points (e.g. mobile base stations, internet access points, etc) may also be fragile. This could be due to electricity load reduction strategies that limit the time grid-based supply is made available (a common occurrence in many other developing countries also). Planning and supporting additional sources of power is essential, and could be in the form of lead-acid batteries, solar panels, etc.

\item Understanding the language in which the questionnaire should be developed can be a challenge, especially if the system needs to be operate in rural areas. Literacy levels must be taken into consideration, along with local dialects which may limit the quality of data that is obtained using such automated approaches. A voice-to-text system may also be used to enable participants to submit their responses in voice format, which can subsequently be converted into text prior to transmission. However, such a system requires an app. to be supported on the device of the participant.

\end{itemize}

\section{Conclusions}
\label{sec:conclusion}

We describe a system to support data collection for carbon credit initiatives, particularly for regions with limited internet coverage. An SMS gateway is proposed that enables an NGO (for instance) to send out a questionnaire to potential participants, and the subsequent responses can then be recorded into a database for further analysis. Such a system may be used directly by participants involved in the carbon credit scheme (depending on the levels of literacy within a region) to volunteers who act as intermediaries to collect the data. The proposed approach can significant increase the frequency a which data collection can be carried out, to supporting reporting of findings to potential donors and investors.

In this instance, a Web-based visual interface is provided for the NGO to: (i) create a repository of questions that can be re-used across questionnaires. Such a repository is particularly useful if multiple languages (or dialects) exist within a particular region, enabling an administrator to choose questions appropriately; (ii) create multiple questionnaires that can be sent out to different participant groups; (iii) create analysis queries on the collected data, the results of which can be visualised using a \textit{word cloud} -- to highlight common terms in the collected responses, or to analyse the data based on demographic information. The data can also be exported to a third party system, such as Matlab or SPSS, for further analysis. The current system uses a central database to hold questions and the results of analysis. This can also be extended to a cloud-hosted database that could be accessible by an NGO that operates in mutiple regions. The data flow presented in Figure~\ref{data_flow} can be extended with a cloud-based system, where the Web application and database can be scaled on-demand.

In addition to the general purpose approach for supporting data collection, we also discuss how this particular system has been used by Vita Green Impact Fund in Eritrea. We investigated a number of factors that could limit potential take up of the proposed system in practice, and limitations with practical deployment of the system due to infrastructural limitations (telecomm. network \& energy grid).

\bibliographystyle{ACM-Reference-Format}

\bibliography{references_ictd}

\end{document}